\documentclass[sigconf]{acmart}

\copyrightyear{2026}
\acmYear{2026}
\setcopyright{cc}
\setcctype{by}
\acmConference[WWW '26]{Proceedings of the ACM Web Conference 2026}{April 13--17, 2026}{Dubai, United Arab Emirates}
\acmBooktitle{Proceedings of the ACM Web Conference 2026 (WWW '26), April 13--17, 2026, Dubai, United Arab Emirates}
\acmPrice{}
\acmDOI{10.1145/3774904.3792734}
\acmISBN{979-8-4007-2307-0/2026/04}

\usepackage{hyperref}
\hypersetup{
    colorlinks=true,
    linkcolor=blue,
    citecolor=blue,
    urlcolor=red
}
\usepackage{bbm}
\usepackage{subfigure}
\usepackage{multirow}
\usepackage{bm}
\usepackage{xcolor}
\usepackage{color, soul}
\definecolor{codegray}{rgb}{0.5,0.5,0.5}

\usepackage{multirow}
\usepackage{enumitem}
\usepackage{graphicx}
\usepackage{algorithmic}
\usepackage[ruled,vlined,linesnumbered]{algorithm2e}
\usepackage{enumitem}

\usepackage[normalem]{ulem}
\useunder{\uline}{\ul}{}
\usepackage{makecell}
\usepackage{etoc}
\etocdepthtag.toc{mtchapter}
\etocsettagdepth{mtchapter}{subsection}
\etocsettagdepth{mtappendix}{none}

\begin{document}

\title{Relational Database Distillation: From Structured Tables to Condensed Graph Data}

\author{Xinyi Gao}
\affiliation{
\institution{The University of Queensland}
\city{Brisbane}
\country{Australia}
}
\email{xinyi.gao@uq.edu.au}

\author{Jingxi Zhang}
\affiliation{
\institution{The University of Queensland}
\city{Brisbane}
\country{Australia}
}
\email{jingxi.zhang@student.uq.edu.au}

\author{Lijian Chen}
\affiliation{
\institution{The University of Queensland}
\city{Brisbane}
\country{Australia}
}
\email{lijian.chen@uq.edu.au}

\author{Tong Chen}
\affiliation{
\institution{The University of Queensland}
\city{Brisbane}
\country{Australia}
}
\email{tong.chen@uq.edu.au}

\author{Lizhen Cui}
\affiliation{
\institution{Shandong University}  
\city{Jinan}
\country{China}
}
\email{clz@sdu.edu.cn}

\author{Hongzhi Yin}
\authornote{Corresponding author.}
\affiliation{
\institution{The University of Queensland}  
\city{Brisbane}
\country{Australia}
}
\email{h.yin1@uq.edu.au}

\renewcommand{\shortauthors}{Xinyi Gao et al.}

\begin{abstract}
Relational databases (RDBs) underpin the majority of global data management systems, where information is structured into multiple interdependent tables. In social media platforms, for instance, massive user-generated data are organized across related tables such as users, posts, comments, and interactions, enabling large-scale analysis and predictive modeling of social behaviors. To effectively use the knowledge within RDBs for predictive tasks, recent advances leverage graph representation learning to capture complex inter-table relations as multi-hop dependencies. Despite achieving state-of-the-art performance, these methods remain hindered by the prohibitive storage overhead and excessive training time, due to the massive scale of database and the computational burden of intensive message passing across interconnected tables. To alleviate these concerns, we propose and study the problem of Relational Database Distillation (RDD). Specifically, we aim to distill large-scale RDBs into compact heterogeneous graphs while retaining the predictive power (i.e., utility) required for training graph-based models. Multi-modal column information is preserved through node features, and primary–foreign key relations are encoded via heterogeneous edges, thereby maintaining both data fidelity and relational structure. To ensure adaptability across diverse downstream tasks without engaging the traditional, inefficient bi-level distillation framework, we further design a kernel ridge regression-guided objective with pseudo-labels, which produces quality features for the distilled graph. Extensive experiments on multiple real-world RDBs demonstrate that our solution substantially reduces the data size while maintaining competitive performances on classification and regression tasks, creating an effective pathway for scalable learning with RDBs.
\end{abstract}

\ccsdesc[500]{Computing methodologies~Neural networks}

\keywords{Relational database, Dataset distillation, Multi-tasks, Efficiency.}

\maketitle

\section{Introduction}
\label{sec_intro}

Relational databases (RDBs) are the foundation of global data management systems \cite{jatana2012survey} and estimated to store over 70\% of the world’s structured data \cite{dbengines2023}. By organizing information into multiple interlinked tables connected through primary–foreign key relations, RDBs capture both rich structural dependencies and diverse statistical patterns. This generalizable framework serves as the foundational backbone for data storage across critical domains, including finance, education, and e-commerce \cite{ilyas2008survey, yin2016spatio,wang2025id,yin2025device,wang2024unveiling}. For example, in social media platforms \cite{yin2015dynamic,hung2017computing}, the relational paradigm models users, posts, comments, and interactions as interdependent entities distributed across multiple tables. This enables a wide range of predictive tasks such as user behavior modeling, recommendation and influence estimation, demonstrating the crucial role of RDBs in understanding large-scale systems.

With the rapid growth of data-driven services, RDBs have expanded to unprecedented scales, often containing billions of rows distributed across interdependent tables. While this growth underscores their central role in data-centric applications, it also introduces significant challenges for predictive modeling \cite{vogel2024wikidbs,tran2025device,tran2025thorough}, particularly in capturing the heterogeneous table attributes and inter-table dependencies induced by foreign-key relations.
In an influence prediction task on social media platforms, for example, modeling user influence requires integrating information across multiple tables. The Users table provides profile and demographic attributes, the Posts and Comments tables contain behavioral and content features, and the Interactions table captures the network structure through user identifiers. Effectively combining these heterogeneous attributes and inter-table dependencies is crucial for accurate influence estimation. However, traditional tabular learning pipelines \cite{fey2024position} rely on extensive feature engineering to join and aggregate related table entities. This process is too labor-intensive to scale and often sacrifices coverage of multi-hop relational information that provides valuable predictive signals.

To alleviate the intricate inter-table dependencies, recent research \cite{zhang2023gfs,robinson2024relbench,cvitkovic2020supervised, dwivedi2025relational,chen2025relgnn, li2025graph} has introduced a promising paradigm that reformulates relational databases as heterogeneous graphs \cite{hu2020heterogeneous}. In this paradigm, table rows/entities are treated as nodes and primary–foreign key relations are modeled as edges. This reformulation enables heterogeneous graph neural networks (GNNs) \cite{liu2025teaching,wang2025graph,wang2025graph,gao2023semantic} to capture inter-table dependencies and multi-hop relations through message passing \cite{gilmer2017neural}, thereby producing expressive node representations for downstream tasks. Benefiting from graph representation learning, these methods eliminate the need for manual feature engineering and have achieved state-of-the-art performances.

Despite its potential to capture relational information, scaling GNNs to large RDBs remains a formidable challenge. GNNs rely on iterative message passing, which requires repeatedly fetching and aggregating neighboring information across multiple interconnected tables. In large-scale RDBs, this leads to prohibitive training time and high computational cost, compounded by substantial storage overhead. As a result, while GNNs advance predictive performance, their scalability barriers limit practical deployment in resource-constrained or time-sensitive environments \cite{gao2024accelerating,gao2024graph1}.

To address these limitations, we investigate Relational Database Distillation (RDD), a new data-centric paradigm that compresses a large relational database into a compact representation while preserving its essential predictive information. The synthetic data enables much more efficient model training yet achieves performance comparable to training on the original large-scale database. While dataset distillation \cite{wang2018dataset,gao2025rethinking,gao2025contrastive} was first explored in computer vision, synthesizing compact data for relational databases introduces several unique challenges due to the nature of tables and relational structures \cite{gao2025robgc,gao2024graph}.
\textit{(i) Multi-modal attributes within tables}. Columns in relational tables are often heterogeneous in modality \cite{cvitkovic2020supervised}, ranging from numerical and categorical to temporal values, and they are inherently unordered. Consequently, each modality necessitates dedicated encoding and distillation strategies to preserve the representativeness of the compressed data and ensure its utility for model training and downstream tasks.
\textit{(ii) Heterogeneous tables and complex structural dependencies}. 
RDBs consist of heterogeneous tables linked by foreign-key relations \cite{fey2024position}, where child tables depend on parent tables. This hierarchical dependency compels conventional generation methods \cite{SDV} to construct parent tables before child tables, thereby increasing modeling complexity and incurring substantial generation overhead. Preserving such dependencies in a compressed form is particularly challenging, as it requires capturing both the heterogeneity of table attributes and the inter-table relationships.
\textit{(iii) Diverse downstream tasks}. Due to their extensive industrial adoption, relational databases are expected to support diverse predictive tasks across a broad range of applications \cite{robinson2024relbench}. While dataset distillation has been explored in modalities such as images \cite{cazenavette2022dataset}, videos \cite{wang2024dancing} and graphs \cite{gao2024graphsurvey}, existing methods are predominantly designed for classification and fail to generalize to other tasks (e.g., regression). Therefore, in various downstream tasks, an effective RDD approach is expected to maintain consistent utility of the synthetic data.

In light of these challenges, we propose Table-to-Graph (T2G), the first relational database distillation framework that compresses large-scale relational databases into compact heterogeneous graphs while preserving predictive utility.
T2G addresses the challenge of multi-modal attributes through a clustering-based pretraining objective. It independently encodes each column using lightweight, modality-specific tokenizers, which can be directly applied for table representation and inference during deployment.
The clustering process further identifies shared patterns among entities and assigns pseudo-labels that group similar ones, forming the foundation for effective distillation.
Subsequently, T2G generates the synthetic heterogeneous graph structure via Stochastic Block Modeling (SBM) \cite{lee2019review} guided by pseudo-labels, which explicitly captures inter-cluster relationships and preserves the structural dependencies across tables.
This design enables the simultaneous generation of diverse edge types, improving efficiency by avoiding sequential table reconstruction.
Finally, to enhance adaptability across tasks, T2G employs a kernel ridge regression (KRR) objective \cite{nguyen2021dataset} for feature distillation. Guided by both task labels and pseudo-labels, the distillation process enhances data quality and ensures adaptability for both classification and regression tasks.

The main contributions of this paper are three-fold:
\begin{itemize}[leftmargin=*]
\item We introduce and formalize relational database distillation (RDD), a novel, practical, yet challenging research problem in large-scale RDB modeling.
\item We propose Table-to-Graph (T2G), a novel RDD framework that distills relational databases into compact heterogeneous graphs. T2G employs a clustering-based objective to independently condense multi-modal attributes and leverages SBM to capture heterogeneous dependencies across tables. Furthermore, it generalizes to both classification and regression tasks, ensuring applicability in diverse predictive scenarios.
\item Through extensive experiments on real-world relational databases, we show that T2G significantly reduces data size while maintaining competitive performances on classification and regression tasks, demonstrating its effectiveness as a scalable paradigm for learning on relational databases. Our codes are available at: \href{https://github.com/XYGaoG/T2G}{https://github.com/XYGaoG/T2G}
\end{itemize}

\section{Preliminaries}
In this section, we first revisit the fundamental concepts of RDBs. 
Subsequently, we show how to incorporate the graph representation learning in the prediction of RDBs.
Finally, we formally define the problem we studied.

\subsection{Relational Database}

A relational database contains a collection of tables $\mathcal{T} = \{T_1, \ldots, T_{n}\}$ and a set of links between them $\mathcal{L} \subseteq \mathcal{T} \times \mathcal{T}$.  
Each table $T \in \mathcal{T}$ is defined as a set of entities/rows $T = \{v_1, \ldots, v_{n_T}\}$, where each element $v_i \in T$ represents a entity/row.  
A link $L = (T_{\mathrm{fkey}}, T_{\mathrm{pkey}}) \in \mathcal{L}$ exists if a foreign key column in table $T_{\mathrm{fkey}}$ references a primary key column in table $T_{\mathrm{pkey}}$. 

Each entity $v \in T$ is represented as a triplet $v = (p_v, \mathcal{K}_v, \textbf{x}_v)$.  
Specifically, $p_v$ denotes the \textbf{primary key} that uniquely identifies the entity $v$.  
$\mathcal{K}_v \subseteq \{p_{v'} : v' \in T' \wedge (T, T') \in \mathcal{L}\}$ represents the set of \textbf{foreign keys}, which define the links from entity $v \in T$ to entities $v' \in T'$, where $p_{v'}$ is the primary key of $v'$ in table $T'$.  
An entity can contain one or more foreign keys, depending on the schema design.  
Finally, $\textbf{x}_v$ denotes the \textbf{multi-modal attributes}, which include numerical and categorical features \footnote{We follow previous works \cite{jurkovicsyntherela, ketata2025joint} to treat the temporal features as the numerical features.}.  
Formally, $\textbf{x}_v$ is a tuple of attribute values: $\textbf{x}_v = (\textbf{x}_v^{\text{num}}, \textbf{x}_v^{\text{cat}})$, where $\textbf{x}_v^{\text{num}} = (x_{v,1}^{\text{num}}, \ldots, x_{v,d_{\text{num}}}^{\text{num}})  \in \mathbb{R}^{d_{\text{num}}}$ denotes the numerical attributes. $\textbf{x}_v^{\text{cat}} = (x_{v,1}^{\text{cat}}, \ldots, x_{v,d_{\text{cat}}}^{\text{cat}})$ denotes the categorical attributes and each $x_{v,i}^{\text{cat}} \in \{1, \ldots, C_i\}$ denotes the discrete category index of the $i$-th attribute.

All primary keys, foreign keys, and attributes are represented as table columns, and all entities within the same table share the same column schema.

In predictive applications, one specific table serves as the \textbf{target type}, denoted as $T_{\text{tgt}} \in \mathcal{T}$.  
The entities in $T_{\text{tgt}}$ are associated with \textbf{task labels} 
$\mathbf{y}_{T_{\text{tgt}}}$, which provide supervision for downstream learning.  
The remaining tables $\mathcal{T} \setminus \{T_{\text{tgt}}\}$ act as auxiliary sources of information that provide contextual signals through relational links, enabling joint reasoning over the entire RDB.

\begin{figure*}[ht]
\centering
\includegraphics[width=\linewidth]{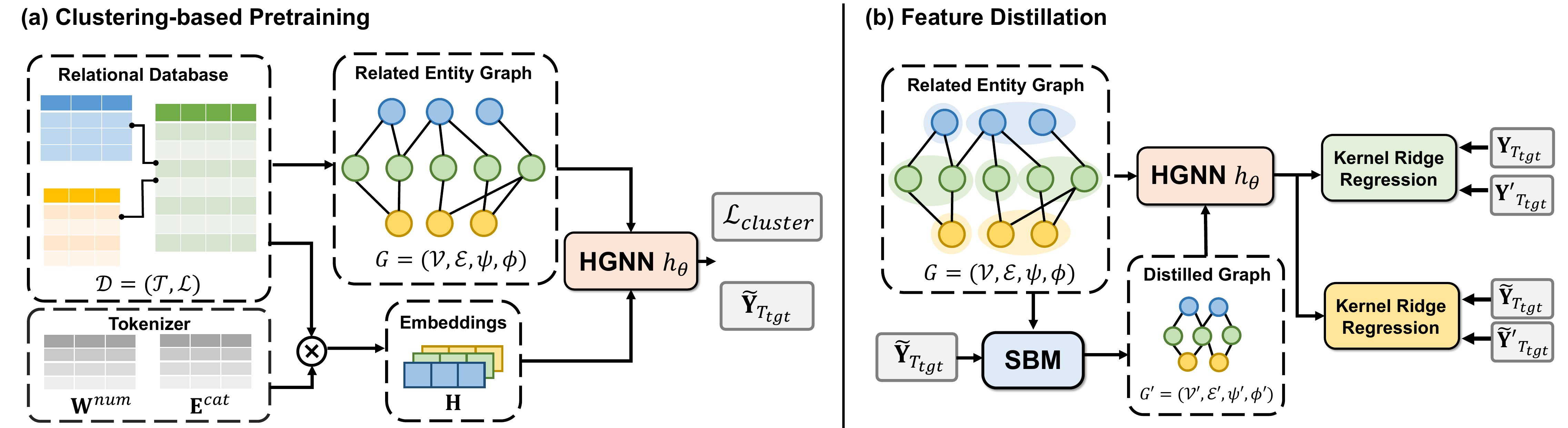}
\caption{The pretraining and distillation procedures of T2G.}
\label{mainfig}
\end{figure*}

\subsection{Graph Modeling of RDB}

The recent advances \cite{dwivedi2025relational,cvitkovic2020supervised, fey2024position}, including relational deep learning (RDL) \cite{robinson2024relbench}, represents each relational database as a heterogeneous graph, referred to as the \textbf{Relational Entity Graph} (REG).  
Given a relational database $(\mathcal{T}, \mathcal{L})$, these methods first augment the link set by including its inverse: 
$\mathcal{L}^{-1} = \{(T_{\text{pkey}}, T_{\text{fkey}}) \mid (T_{\text{fkey}}, T_{\text{pkey}}) \in \mathcal{L}\}$.  
Accordingly, the schema graph is defined as $(\mathcal{T}, \mathcal{R})$, where the node set is $\mathcal{T}$ and the edge set is $\mathcal{R} = \mathcal{L} \cup \mathcal{L}^{-1}$. This describes the table-level structure of data and indicates type definitions.
Based on this schema, the REG is defined as a heterogeneous graph where each row in table is represented by a node. REG is denoted as $G = (\mathcal{V}, \mathcal{E}, \phi, \psi)$, where $\mathcal{V}$ is the node set, $\mathcal{E} \subseteq \mathcal{V} \times \mathcal{V}$ is the edge set, $\phi : \mathcal{V} \to \mathcal{T}$ is the node-type mapping function, and $\psi : \mathcal{E} \to \mathcal{R}$ is the edge-type mapping function.  
Each node $v \in \mathcal{V}$ is assigned a type $\phi(v) \in \mathcal{T}$, and nodes of the same type correspond to entities in the same table.  
Similarly, each edge $e \in \mathcal{E}$ is assigned a type $\psi(e) \in \mathcal{R}$ as specified by the schema.  

By transforming relational databases into REGs, heterogeneous graph neural networks (GNNs) can be applied to encode entities and generate representations for downstream prediction tasks via message passing.  
The multi-modal input attributes $\textbf{x}_v $ are first processed by modality-specific encoders \cite{fey2024position}, producing embeddings for each column that are then fused into a unified entity representation $\{\mathbf{h}_v^{(0)}\}_{v \in \mathcal{V}}$.  
Subsequently, a heterogeneous GNN iteratively updates the embeddings according to neighboring entities:
\begin{equation}
\label{HGNN}
\footnotesize
\mathbf{h}_v^{(i+1)} = f_{\phi(v)}\Bigg(
    \mathbf{h}_v^{(i)}, 
    \Big\{\!\!\Big\{ 
        f_R\big(\{\!\!\{ g_R(\mathbf{h}_w^{(i)}) \mid w \in \mathcal{N}_R(v) \}\!\!\}\big) 
        \;\Big|\; \forall R = (T, \phi(v)) \in \mathcal{R} 
    \Big\}\!\!\Big\} 
\Bigg),
\end{equation}
where $f_{\phi(v)}$, $f_R$, and $g_R$ are differentiable transformation functions, and $\{\!\!\{\cdot\}\!\!\}$ denotes an aggregation operator such as mean, max, or sum.  
$\mathcal{N}_R(v) = \{ w \in \mathcal{V} \mid (w,v) \in \mathcal{E}, \, \psi(w,v) = R \}$ represents the $R$-specific neighborhood of node $v$.  

After multiple propagation layers, the final embeddings integrate information from multi-hop heterogeneous neighbors. The embeddings for the target type $T_{\text{tgt}}$ are then fed into task-specific prediction heads for downstream applications such as classification or regression.

\subsection{Problem Formulation}
We consider a relational database $\mathcal{D} = (\mathcal{T}, \mathcal{L})$, and its schema graph $(\mathcal{T}, \mathcal{R})$. The corresponding  REG is $G = (\mathcal{V}, \mathcal{E}, \phi, \psi)$.
The entities in $T_{\text{tgt}}$ are associated with task labels $\mathbf{y}_{T_{\text{tgt}}}$.

Relational Database Distillation (RDD) aims to construct a compact synthetic database  
$\mathcal{D}' = (\mathcal{T}', \mathcal{L}')$ as well as the task label $\mathbf{y}'_{T_{\text{tgt}}}$, such that the total number of entities in $\mathcal{T}'$  
is substantially smaller than in $\mathcal{T}$, while preserving the predictive information contained in $\mathcal{D}$.  

In this work, we reparameterize $\mathcal{D}'$ as the tuple $\mathcal{D}' = (\mathcal{M}, G')$,  
where $\mathcal{M} = \{M_1, \ldots, M_m\}$ is a set of lightweight column tokenizers and  
each $M_i$ encodes column attributes into compact embeddings.  
$G' = (\mathcal{V}', \mathcal{E}', \phi', \psi')$ is the synthetic REG,  
where $\mathcal{V}'$ is the set of synthetic nodes,  
$\mathcal{E}' \subseteq \mathcal{V}' \times \mathcal{V}'$ the set of edges,  
$\phi': \mathcal{V}' \to \mathcal{T}'$ assigns each node to a table type,  
and $\psi': \mathcal{E}' \to \mathcal{L}'$ assigns each edge to a relation type. The target type entities/nodes are associated with the task label $\mathbf{y}'_{T_{\text{tgt}}}$. $|\mathcal{V}'|\ll|\mathcal{V}|$ and the schema of the synthetic REG is required to match that of the original database. With this formulation, heterogeneous GNN models trained on ${G}'$  
are expected to achieve performance on downstream tasks comparable to those trained on the original $G$, while requiring significantly less storage and training time. 

In the inference stage, the test REG is first constructed from the RDB, and the pretrained tokenizers $\mathcal{M}$ are employed to independently encode each column. Then, the trained GNN model is applied to perform prediction according to the encoded REG and embeddings.

\section{Methodologies}
We hereby present Table-to-Graph (T2G), a relational database distillation framework that compresses large-scale relational databases into compact heterogeneous graphs. As shown in Figure~\ref{mainfig}, T2G begins by encoding each table using lightweight, modality-specific tokenizers pretrained with a clustering-based objective, which captures common patterns across entities and produces pseudo-labels through cluster assignments. Guided by these pseudo-labels, T2G employs SBM to reconstruct heterogeneous graph structures and generate a compact REG that maintains the structural dependencies. Finally, entity features in the synthetic REG are distilled through a KRR objective supervised by both task labels and pseudo-labels, retaining task-relevant information and supporting efficient model training.

\subsection{Modality-specific Tokenizers}

RDBs contain multiple tables, each comprising diverse multi-modal attributes.  
For clarity, we illustrate the column encoding process using a single table in this section.  
Formally, each entity $v \in \mathcal{V}$ is characterized by a set of attributes 
$\mathbf{x}_v = (\mathbf{x}_v^{\text{num}}, \mathbf{x}_v^{\text{cat}})$, 
where $\textbf{x}_v^{\text{num}} = (x_{v,1}^{\text{num}}, \ldots, x_{v,d_{\text{num}}}^{\text{num}})  \in \mathbb{R}^{d_{\text{num}}}$ denotes the numerical attributes. $\textbf{x}_v^{\text{cat}} = (x_{v,1}^{\text{cat}}, \ldots, x_{v,d_{\text{cat}}}^{\text{cat}})$ denotes the categorical attributes and each $x_{v,i}^{\text{cat}} \in \{1, \ldots, C_i\}$ denotes the discrete category index.
 
To effectively encode these heterogeneous modalities, T2G employs two types of lightweight, modality-specific tokenizers that independently process each column.

For numerical attributes, each column value $x_{v,i}^{\text{num}}$ is projected into the embedding space using a learnable tokenizer $\mathbf{W}_{i}^{\text{num}} \in \mathbb{R}^{1 \times d_{\text{token}}}$:
\begin{equation}
\mathbf{h}_{v,i}^{\text{num}} = x_{v,i}^{\text{num}} \mathbf{W}_{i}^{\text{num}}, \quad i = 1, \ldots, d_{\text{num}}.
\end{equation}
This produces column-wise embeddings 
$\mathbf{H}_v^{\text{num}} = [\mathbf{h}_{v,1}^{\text{num}}, \ldots, \mathbf{h}_{v,d_{\text{num}}}^{\text{num}}] \in \mathbb{R}^{d_{\text{num}} \times d_{\text{token}}}$.

For categorical attributes, each column $x_{v,i}^{\text{cat}}$ takes values from a discrete category set $\{1, \ldots, C_i\}$ and is encoded using a learnable lookup table $\mathbf{E}_{i}^{\text{cat}} \in \mathbb{R}^{C_i \times d_{\text{token}}}$.  
The embedding for the $i$-th categorical column is retrieved as:
\begin{equation}
\mathbf{h}_{v,i}^{\text{cat}} = \mathbf{E}_{i}^{\text{cat}}[x_{v,i}^{\text{cat}}], \quad i = 1, \ldots, d_{\text{cat}},
\end{equation}
resulting in the embeddings
$\mathbf{H}_v^{\text{cat}} = [\mathbf{h}_{v,1}^{\text{cat}}, \ldots, \mathbf{h}_{v,d_{\text{cat}}}^{\text{cat}}] \in \mathbb{R}^{d_{\text{cat}} \times d_{\text{token}}}$.

Each attribute is thus represented by a dedicated tokenizer and embedded into a $d_{\text{token}}$-dimensional space.  
The complete tokenizer parameters consist of 
$\mathbf{W}^{\text{num}}\in \mathbb{R}^{d_{\text{num}} \times d_{\text{token}}}$ and 
$\mathbf{E}^{\text{cat}}  \in \mathbb{R}^{C \times d_{\text{token}}}$, where $C$ denotes the total number of categories in the table.
Notably, $d_{\text{num}} + C$ is proportional to the number of columns in the table and is typically much smaller than the number of entities. Moreover, $d_{\text{token}}$ can be set to a small value (i.e., 4 to 8 in practice) while maintaining sufficient representational capacity, ensuring these tokenizers are memory-efficient during deployment.

Importantly, each tokenizer can be viewed as a distilled representation of its corresponding column:  
the original column values serve as \textit{keys} to retrieve the embeddings from the tokenizer, effectively replacing the raw data with a lightweight yet informative mapping.  
Thus, storing $\mathcal{M} = \{\mathbf{W}^{\text{num}}, \mathbf{E}^{\text{cat}}\}$ captures the essential column-level knowledge of the database, enabling efficient storage, transfer, and inference while preserving task-relevant information. We empirically justify the compact size of the tokenizers through experiments presented in Section~\ref{sec_distilldata}.

To obtain a unified entity representation, the embeddings from both modalities are aggregated through an aggregation function $\operatorname{AGG}(\cdot)$, which can be instantiated as mean, sum, or max pooling:
\begin{equation}
\mathbf{h}_v^{(0)} = \operatorname{AGG}\big(\mathbf{H}_v^{\text{num}}, \mathbf{H}_v^{\text{cat}}\big),
\end{equation}
where $\mathbf{h}_v^{(0)} \in \mathbb{R}^{d_{\text{token}}}$ serves as the initial embedding of entity $v$.

\subsection{Clustering-based Pre-training}

After obtaining the embeddings for all entities, the encoded representations of each table are organized into embedding matrices 
$\mathbf{H}_{T} = [\mathbf{h}_{v_1}^{(0)}, \ldots, \mathbf{h}_{v_{n_T}}^{(0)}] \in \mathbb{R}^{n_T \times d_{\text{token}}}$, 
where $n_T$ is the number of entities in table $T \in \mathcal{T}$. 
These embeddings $\mathbf{H} = \{\mathbf{H}_T \mid T \in \mathcal{T}\}$, along with the REG $G$, are then processed by a heterogeneous graph neural network (HGNN) $h_\theta$ as defined in Eq.~(\ref{HGNN}):
\begin{equation}
\mathbf{Z} = h_\theta(G, \mathbf{H}),
\end{equation}
where $\mathbf{Z} = \{\mathbf{Z}_T \mid T \in \mathcal{T}\}$. The HGNN captures inter-table dependencies by propagating information along primary--foreign key relations and outputs processed embeddings 
$\mathbf{Z}_T = [\mathbf{z}_{v_1}, \ldots, \mathbf{z}_{v_{n_T}}]$ for each table.

To guide representation learning, we employ a clustering-based pre-training objective that encourages embeddings to form semantically coherent groups. 
For each table $T$, online clustering is performed on $\mathbf{Z}_T$ to obtain pseudo-labels 
$\tilde{\mathbf{y}}_T = \{\tilde{y}_{v_1}, \ldots, \tilde{y}_{v_{n_T}}\}$, where each $\tilde{y}_{v_i} \in \{1, \ldots, n'_T\}$ indicates the cluster assignment of entity $v_i$. 
Here, $n'_T$ also corresponds to the number of synthetic entities allocated to table $T$ in the synthetic graph $G'$, ensuring a one-to-one correspondence between clusters and synthetic entities.
The optimization follows a classification objective:
\begin{equation}
\label{pretrain}
\mathcal{L}_{\text{cluster}} = - \sum_{T \in \mathcal{T}} \sum_{v \in T} \log P(\tilde{y}_v \mid \mathbf{z}_v),
\end{equation}
which jointly updates the encoder parameters and pseudo-label assignments to discover discriminative and well-separated clusters. 
The resulting pseudo-labels provide the structural foundation for constructing the synthetic REG.

\noindent\textbf{Clustering for target entities.}
To make the clustering process task-aware, we adopt distinct strategies to generate the pseudo-labels for target type entities, instead of unsupervised clustering:
\begin{itemize}[leftmargin=*]
\item For classification tasks, clustering is performed \textit{within each class} of the target entities to avoid semantic mixing between categories. 
Entities sharing the same task label are clustered independently in the embedding space, ensuring that each cluster represents a coherent subpopulation within its class. 
\item For regression tasks, clustering is instead performed directly over the continuous labels, allowing the model to capture local label density and provide more fine-grained supervision information.
\end{itemize}

\subsection{Graph Generation via Heterogeneous SBM}

Based on the pseudo-labels $\tilde{\mathbf{y}}_T$ and clusters identified from the pretraining stage, we next construct the synthetic REG by modeling inter-table dependencies using a heterogeneous Stochastic Block Model (SBM). 
The pseudo-labels partition table $T \in \mathcal{T}$ into $n'_T$ clusters, each of which corresponds to a synthetic entity in the synthetic graph. 
Each cluster thus offers a block of prototypes for the corresponding entities in the synthetic REG $G' = (\mathcal{V}', \mathcal{E}', \phi', \psi')$.

To capture the heterogeneous dependencies across tables, we construct a type-specific connectivity matrix $\mathbf{P}_{T_i,T_j} \in [0,1]^{n'_{T_i} \times n'_{T_j}}$ for each relation $(T_i,T_j) \in \mathcal{R}$. 
Each entry $\mathbf{P}_{T_i,T_j}(a,b)$ models the empirical probability of linkage between entities belonging to cluster $a$ in $T_i$ and cluster $b$ in $T_j$, estimated as:
\begin{equation}
\label{prob}
\mathbf{P}_{T_i,T_j}(a,b) = 
\frac{
\sum_{(u,v)\in\mathcal{E}_{i,j}}  
\mathbb{I}\!\left[\tilde{y}_{v}=a,\, \tilde{y}_{w}=b\right]
}{
\sum_{v \in T_i, w \in T_j}
\mathbb{I}\!\left[\tilde{y}_{v}=a,\, \tilde{y}_{w}=b\right]
}.
\end{equation}
Here, $a \in \{1,\ldots,n'_{T_i}\}$, $b \in \{1,\ldots,n'_{T_j}\}$, and $\mathbb{I}[\cdot]$ denotes the indicator function. 
Intuitively, $\mathbf{P}_{T_i,T_j}(a,b)$ represents the normalized connection density between the two clusters. 
Collectively, the set $\{\mathbf{P}_{T_i,T_j}\}_{(T_i,T_j)\in \mathcal{R}}$ defines a heterogeneous SBM that jointly characterizes the inter-table structural dependencies.

Afterwards, we synthesize the relational structure by applying a deterministic thresholding mechanism to control graph sparsity.
Specifically, a global sparsity ratio $\rho$ is set as a hyper-parameter, and an adaptive threshold $\tau_{i,j}$ is derived for each relation type to preserve its original density:
\begin{equation}
\label{eq:threshold}
\tau_{i,j} = \mathrm{Quantile}_{1-\rho}\!\big(\mathbf{P}_{T_i,T_j}\big),
\end{equation}
where $\mathrm{Quantile}_{1-\rho}$ denotes the $(1-\rho)$-quantile of all entries in $\mathbf{P}_{T_i,T_j}$.
The edges in the adjacency matrix $\mathbf{A}'_{T_i,T_j}$ are then deterministically defined as:
\begin{equation}
\label{eq:edge_gen}
\mathbf{A}'_{T_i,T_j}(a,b) =
\begin{cases}
1, & \text{if } \mathbf{P}_{T_i,T_j}(a,b) > \tau_{i,j},\\
0, & \text{otherwise.}
\end{cases}
\end{equation}
By this means, we can jointly capture relation-specific connectivity patterns across multiple tables. Furthermore, this eliminates the need for hierarchical table generation, enabling parallel feature synthesis across all tables conditioned on the generated graph structure.

\subsection{KRR-based Feature Distillation}

After obtaining the heterogeneous structure $G'$, the final stage of RDD focuses on generating features for the synthetic entities. 
To this end, we employ a HGNN $h_\theta$ to encode both the original and synthetic graphs:
\begin{equation}
\mathbf{Z} = h_\theta(G, \mathbf{H}), 
\quad 
\mathbf{Z}' = h_\theta(G', \mathbf{H}'),
\end{equation}
where $\mathbf{H}' = \{\mathbf{H}'_T \mid T \in \mathcal{T}\}$ denotes the features of synthetic graph and each $\mathbf{H}'_T \in \mathbb{R}^{n'_T \times d_{\text{token}}}$ is initialized randomly with the same dimension as $\mathbf{H}_T \in \mathbb{R}^{n_T \times d_{\text{token}}}$.  

Our goal is to optimize the synthetic embeddings $\mathbf{H}'$ such that a model trained on the synthetic data achieves comparable predictive performance to the one trained on original data.  
However, this process inherently involves a nested optimization problem \cite{yu2023dataset}, where both the predictive model and the synthetic data must be optimized recurrently.  
In single-modality data distillation, such bi-level optimization design has already exposed  significant computational complexity and training instability \cite{wang2018dataset, yu2023dataset}. Thus, straightforwardly adopting bi-level optimization for RDD will further impair the distillation quality, given RDBs' larger scale, multi-modality, and intricate high-order interactions.

To alleviate this issue, we replace the conventional trainable predictor with a \textit{Kernel Ridge Regression} (KRR) model, which provides a closed-form solution for the optimal weights. This enables direct optimization of the synthetic features $\mathbf{H}'$ without iterative model training. Specifically, the KRR prediction model on the synthetic graph is defined as:
\begin{equation}
\mathcal{L}_{\text{KRR}} = 
\|\mathbf{Y}'_{T_{\text{tgt}}} - \mathbf{Z}'_{T_{\text{tgt}}}\mathbf{W}\|^2 + \lambda \|\mathbf{W}\|^2,
\end{equation}
where $\lambda$ is the ridge regularization coefficient, and $\mathbf{W}$ is the learnable weight matrix. $\mathbf{Y}'_{T{\text{tgt}}}$ denotes matrix-form task labels for target entities.
The optimal solution admits a closed form:
\begin{equation}
\label{krr1}
\mathbf{W}^{*} =
\mathbf{Z}'^{\top}_{T_{\text{tgt}}}
\Big(
\mathbf{Z}'_{T_{\text{tgt}}}\mathbf{Z}'^{\top}_{T_{\text{tgt}}} + \lambda \mathbf{I}
\Big)^{-1}
\mathbf{Y}'_{T_{\text{tgt}}}.
\end{equation}

To ensure the synthetic data yield high predictive performance on the original data, we minimize the discrepancy between the predictions of the original data and their task labels:
\begin{equation}
\mathcal{L}_{\text{task}}(\mathbf{H}', \mathbf{Y}'_{T_{\text{tgt}}}) =
\mathbb{E}_{\theta \sim \Theta}
\Big[
\|\mathbf{Y}_{T_{\text{tgt}}} - \mathbf{Z}_{T_{\text{tgt}}}\mathbf{W}^{*}\|^2
\Big],
\label{eq:taskloss}
\end{equation}
where the expectation form over parameter $\theta$ encourages generalization across different model initializations $\Theta$ while KRR guarantees an optimal solution under random feature mappings \cite{loo2022efficient}.  
KRR-based objective accommodates both regression and classification tasks, where categorical labels are converted into one-hot format for computation.

\noindent\textbf{Extension with pseudo-labels.}
To further enhance structural consistency and representation alignment, we introduce an auxiliary KRR loss based on the pseudo-labels $\tilde{\mathbf{Y}}_{T_{\text{tgt}}}$ obtained from the clustering-based pretraining:
\begin{equation}
\mathcal{L}_{\text{pseudo}}(\mathbf{H}', \tilde{\mathbf{Y}}'_{T_{\text{tgt}}}) =
\mathbb{E}_{\theta \sim \Theta}
\Big[
\|\tilde{\mathbf{Y}}_{T_{\text{tgt}}} - \mathbf{Z}_{T_{\text{tgt}}}\mathbf{W}^{*}_{\text{pseudo}}\|^2
\Big],
\label{eq:pseudoloss}
\end{equation}
where 
\begin{equation}
\label{krr2}
\mathbf{W}^{*}_{\text{pseudo}} =
\mathbf{Z}'^{\top}_{T_{\text{tgt}}}
\Big(
\mathbf{Z}'_{T_{\text{tgt}}}\mathbf{Z}'^{\top}_{T_{\text{tgt}}} + \lambda \mathbf{I}
\Big)^{-1}
\tilde{\mathbf{Y}}'_{T_{\text{tgt}}},
\end{equation}
and $\tilde{\mathbf{Y}}'_{T_{\text{tgt}}}$ denotes the identity matrix serving as the pseudo-labels for synthetic data.

\begin{algorithm}[t]
\SetAlgoVlined
\small
\textbf{Input:} Relational database $\mathcal{D} = (\mathcal{T}, \mathcal{L})$; task labels $\mathbf{Y}_{T_{\text{tgt}}}$.\\
\textbf{Output:} Synthetic graph ${G}'$; synthetic features $\mathbf{H}'$; synthetic labels ${\mathbf{Y}}'_{T_{\text{tgt}}}$; tokenizers $\mathbf{W}^{\text{num}}$, $\mathbf{E}^{\text{cat}}$.\\[0.5ex]

{\color{gray}{$\rhd$ \textbf{Stage 1: Clustering-based Pretraining}}}\\
Initialize $\mathbf{W}^{\text{num}}$ and $\mathbf{E}^{\text{cat}}$.\\
\For{$iter = 1, \ldots, K_{\text{pre}}$}{
Encode $T \in \mathcal{T}$ into $\mathbf{H}_T$ by $\mathbf{W}^{\text{num}}$ and $\mathbf{E}^{\text{cat}}$.\\
$\mathbf{Z} = h_\theta(G, \mathbf{H})$.\\
Compute $\mathcal{L}_{\text{cluster}}$ with Eq. (\ref{pretrain}).\\
Update $h_\theta$, $\mathbf{W}^{\text{num}}$ and $\mathbf{E}^{\text{cat}}$.\\
}
Compute $\tilde{\mathbf{Y}}_T$ for $T \in \mathcal{T}$.\\

{\color{gray}{$\rhd$ \textbf{Stage 2: Graph Structure Distillation}}}\\
\For{$(T_i, T_j) \in \mathcal{R}$}{
    Compute $\mathbf{P}_{T_i, T_j}$ using $\tilde{\mathbf{Y}}_{T_{i}}$ and $\tilde{\mathbf{Y}}_{T_{j}}$.\\
    Compute $\mathbf{A}'_{T_i,T_j}$ with sparsity $\rho$.\\
}
Compute $G' = (\mathcal{V}', \mathcal{E}', \phi', \psi')$.\\

{\color{gray}{$\rhd$ \textbf{Stage 3: KRR-based Feature Distillation}}}\\
Initialize $\mathbf{H}'$ and ${\mathbf{Y}}'_{T_{\text{tgt}}}$.\\
\For{$iter = 1, \ldots, K_{\text{distill}}$}{
    Sample $\theta \sim \Theta$ and initialize $h_\theta$.\\
    $\mathbf{Z} = h_\theta(G, \mathbf{H}); \; \mathbf{Z}' = h_\theta(G', \mathbf{H}')$.\\
    Compute KRR model $\mathbf{W}^{*}$ and $\mathbf{W}^{*}_{\text{pseudo}}$  with Eq. (\ref{krr1}) and Eq. (\ref{krr2}).\\
    Compute $\mathcal{L}_{\text{total}}$ with Eq. (\ref{totalloss}).\\
    Update $\mathbf{H}'$ and ${\mathbf{Y}}'_{T_{\text{tgt}}}$.\\
}
\caption{The Procedure of Table-to-Graph (T2G)}
\label{al}
\end{algorithm}

The overall optimization objective combines task supervision and pseudo-label guidance:
\begin{equation}
\mathcal{L}_{\text{total}}(\mathbf{H}', \mathbf{Y}'_{T_{\text{tgt}}}, \tilde{\mathbf{Y}}'_{T_{\text{tgt}}}) =
\mathcal{L}_{\text{task}}(\mathbf{H}', \mathbf{Y}'_{T_{\text{tgt}}}) +
\beta \, \mathcal{L}_{\text{pseudo}}(\mathbf{H}', \tilde{\mathbf{Y}}'_{T_{\text{tgt}}}),
\label{totalloss}
\end{equation}
where $\beta$ balances the contribution of the pseudo-label loss. The optimal features and labels are derived from:
\begin{equation}
\mathbf{H}'^{*}, \mathbf{Y}'^{*}_{T_{\text{tgt}}} = 
\arg\min_{\mathbf{H}', \mathbf{Y}'_{T_{\text{tgt}}}} \mathcal{L}_{\text{total}}.
\end{equation}
This dual-supervision design ensures that the synthetic features and labels encode both task-specific semantics and relational structure, leading to a compact yet expressive synthetic data suitable for diverse downstream applications.

\subsection{Further Detailed Analysis}
\setlength{\abovecaptionskip}{1pt}

\noindent\textbf{Algorithm.} 
The detailed process of T2G is presented in Algorithm~\ref{al}. Initially, we pre-train the tokenizers and generate pseudo-labels for each entity (lines 4–10). These pseudo-labels are then used to guide the construction of the SBM, which in turn generates the heterogeneous graph for the synthetic graph. In lines 17–23, the KRR-based objective is computed to optimize the synthetic features and labels, incorporating both task labels and pseudo-labels of original data.

\noindent\textbf{Model training and deployment.} 
During downstream model training, the synthetic graph structure $G'$, features $\mathbf{H}'$, and labels $\mathbf{Y}'_{T_{\text{tgt}}}$ are used for efficient model optimization. At inference time, the lightweight tokenizers $\mathcal{M}$ are first applied to obtain column and entity embeddings, after which the trained model is used to perform the prediction task.

The detailed time complexity is provided in the Appendix \ref{Tcom}.

\section{Experiments}
\label{sec_exp}

We design comprehensive experiments to validate the efficacy of our proposed methods and explore the following research questions:
\noindent\textbf{Q1}: Compared to the other dataset reduction and distillation methods, can T2G achieve better performance?
\noindent\textbf{Q2}: Can the synthetic data generated by T2G generalize well to different model architectures? 
\noindent\textbf{Q3}: How do the different components, i.e., the SBM module, loss terms, and pseudo-labels, influence the performance of T2G?
\noindent\textbf{Q4}: How do the hyperparameters (the weight $\beta$ and the sparsity $\rho$) affect the behavior of T2G?
\noindent\textbf{Q5}: How effectively does T2G reduce storage size compared to the original database?

The results for Q5 are deferred to Appendix \ref{sec_distilldata}.

\subsection{Experimental Setup}
{\bf Databases \& Baselines.}
We evaluate our proposed methods on 4 real-world databases: Rossmann, Walmart, Airbnb and IMDB, using the same split method as in the RDBs benchmark \cite{jurkovicsyntherela}. These databases differ in the types of relationships, as well as in the number of tables and columns. A summary of key characteristics is provided in Table~\ref{tab:databases}.

\begin{table}[t]
\setlength{\abovecaptionskip}{5pt}
\renewcommand{\arraystretch}{1.2}
\centering
\caption{Summary of Databases. Reg. and Cla. denote the regression and classification tasks, respectively.}
\resizebox{\linewidth}{!}{
\begin{tabular}{lccccc}
\Xhline{1.pt}
{Database} & {\#Rows} & {\#Columns} & {\#Tables} & {Train/Val/Test} & {Task} \\
\hline
Rossmann & 1,015,159 & 16 & 2 & 637,954/177,620/201,815 & Reg. \\
Walmart & 310,707 & 17 & 3 & 154,759/77,679/78,359 & Reg. \\
Airbnb & 1,020,723 & 20 & 2 & 608,911/207,429/204,383 & Cla. \\
IMDB & 2,377,044 & 19 & 6 & 1,133,141/623,680/620,223 & Cla. \\
\Xhline{1.pt}
\end{tabular}
}
\label{tab:databases}
\end{table}

We evaluate the proposed T2G against a diverse set of representative dataset distillation and sampling baselines. 
Specifically, we compare with the following methods:  
(1) {Random}: randomly selects a subset of entities.  
(2) {Herding}~\cite{welling2009herding}: selects representative entities that are closest to the cluster center.  
(3) {K-Center}~\cite{farahani2009facility}: chooses entities that maximize coverage in the feature space by minimizing the maximum pairwise distance.  
(4) {Coarsening}~\cite{huang2021scaling}: constructs a smaller graph by merging similar entities while preserving topological properties.
(5) {FreeHGC}~\cite{liang2025training}: a training-free dataset distillation method for heterogeneous graphs that selects entities according to influence metrics.  
(6) {HGCond}~\cite{gao2024heterogeneous}: a dataset distillation method designed for heterogeneous graphs, which matches model gradients between the original and synthetic data.  

Since these methods are not originally designed for tabular or RDBs, they do not consider column-wise encoding. 
To ensure a fair comparison, we apply all baselines on top of the encoded entity embeddings obtained by pretrained tokenizers, which are supervised with task labels.

\begin{table*}[t]
\renewcommand{\arraystretch}{1.1}
\centering
\caption{
Comparison of model performance trained with different dataset reduction methods. Values are reported as mean ± standard deviation. ``Whole Database'' denotes the model trained on the full original RDB. $r$ is the compression ratio. Rossmann and Walmart are regression tasks evaluated by MAE, while Airbnb and IMDB are classification tasks evaluated by AUC (\%).
}
\label{tab:main}
\resizebox{0.9\linewidth}{!}{%
\begin{tabular}{c|c|cccccc|c|c}
\Xhline{1.pt}
\multirow{2}{*}{Database}              & \multirow{2}{*}{$r$} & \multicolumn{6}{c|}{Baselines}                                           & Ours           & \multirow{2}{*}{\begin{tabular}[c]{@{}c@{}}Whole \\ Database\end{tabular}} \\ \cline{3-9}
                                       &                        & Random     & Herding    & K-Center  & Coarsening & FreeHGC   & HGCond    & T2G                &                                                                            \\ \hline
\multirow{3}{*}{Rossmann $\downarrow$} & 0.02\%                   & 381.7±8.3  & 237.0±11.2 & 324.5±9.1 & 250.1±9.1  & 229.9±5.4 & 215.8±2.0 & \textbf{194.4±3.0} & \multirow{3}{*}{188.5±0.9}                                                 \\
                                       & 0.05\%                  & 361.4±8.4  & 232.2±11.2 & 271.4±7.6 & 237.8±6.9  & 218.4±5.9 & 212.1±4.5 & \textbf{194.8±3.1} &                                                                            \\
                                       & 0.10\%                   & 352.9±9.1  & 221.8±13.2 & 261.3±9.7 & 236.0±6.9  & 219.2±6.6 & 213.1±1.4 & \textbf{192.8±4.1} &                                                                            \\ \hline
\multirow{3}{*}{Walmart $\downarrow$}  & 0.10\%                   & 18736±2598 & 16793±1223 & 16967±653 & 15136±799  & 15185±640 & 14012±647 & \textbf{8989±255}  & \multirow{3}{*}{8790±272}                                                  \\
                                       & 0.20\%                   & 16454±1303 & 16142±1054 & 16223±494 & 14325±393  & 14141±719 & 14145±600 & \textbf{9070±342}  &                                                                            \\
                                       & 0.40\%                   & 16086±832  & 15179±991  & 15492±404 & 14140±322  & 14260±883 & 14072±599 & \textbf{8848±282}  &                                                                            \\ \hline
\multirow{3}{*}{Airbnb $\uparrow$}     & 0.02\%                   & 53.6±0.6   & 56.5±0.5   & 54.9±1.5  & 57.1±1.6   & 58.6±0.8  & 62.8±1.2  & \textbf{65.5±0.5}  & \multirow{3}{*}{69.0±0.1}                                                  \\
                                       & 0.05\%                   & 56.2±1.1   & 57.0±1.3   & 57.8±2.0  & 58.9±1.2   & 59.7±1.1  & 62.6±0.7  & \textbf{66.1±0.7}  &                                                                            \\
                                       & 0.10\%                   & 57.5±1.9   & 58.3±1.2   & 57.9±2.2  & 59.0±1.1   & 60.9±1.9  & 62.6±0.6  & \textbf{66.6±0.5}  &                                                                            \\ \hline
\multirow{3}{*}{IMDB $\uparrow$}      & 0.02\%                   & 50.9±0.2 & 52.0±0.1 & 52.3±0.1 & 54.0±0.4 & 56.4±0.8 & 60.4±1.2 & \textbf{65.2±1.2}  & \multirow{3}{*}{68.5±0.9}                                                  \\
                                       & 0.05\%                   & 51.1±0.3 & 52.3±0.5 & 52.0±1.7 & 55.4±0.5 & 57.2±2.3 & 60.3±1.0 & \textbf{64.8±1.7}  &                                                                            \\
                                       & 0.10\%                   & 53.1±0.9 & 54.2±0.3 & 54.8±1.8 & 55.9±2.4 & 58.3±0.1 & 60.8±1.1 & \textbf{65.5±1.3}  &                                                                            \\ \Xhline{1.pt}

\end{tabular}}
\end{table*}

\noindent{\bf Implementations.}
We evaluate three compression ratios ($r = N'/N$) for each database, where $N$ denotes the number of entities in the original \textit{training set} and $N'$ represents the number of synthetic entities. 
The tokenizer embedding dimension $d_{{token}}$ and HGNN hidden dimension $d$ are set to 8 and 64, respectively. 
For HGNN used in both distillation and evaluation stages, we follow prior work~\cite{hudovernik2024relational, fey2024position} by adopting the GraphSAGE~\cite{hamilton2017inductive} architecture and extending it to support heterogeneous REGs.
The regression and classification tasks are evaluated by MAE and AUC score, respectively.

\noindent\textbf{Hyper-parameters.} 
The hyper-parameters are determined through a grid search on the validation set.
Following \cite{hudovernik2024relational}, the HGNN layer is set as the number of tables in the RDBs. The aggregation function used for entity embedding generation is set to the mean.
The loss weight $\beta$ is tuned from the set \{0, 0.1, 1, 10, 20\}, and the sparsity $\rho$ is explored within the range (0, 1). The ridge regularization coefficient $\lambda$ is set as 1e-2.
The learning rate for the distillation process is determined through a search over the set \{0.1, 0.01, 0.001\}. The weight decay is 5e-4. 

\noindent\textbf{Computing Infrastructure.} The codes are written in Python 3.9 and Pytorch 1.12.1. 
The experiments were conducted on a server equipped with Intel(R) Xeon(R) Gold 6326 CPUs at 2.90GHz and NVIDIA GeForce A40 GPUs with 48GB of memory.

\subsection{Effectiveness Comparison (Q1)}

Table~\ref{tab:main} summarizes the results across all databases, compression ratios, and baselines. Our proposed T2G consistently achieves the best performance, demonstrating its strong ability to preserve predictive information while drastically reducing data size.

Traditional selection-based baselines such as Random, Herding, and K-Center exhibit unstable results. 
In particular, K-Center heavily depends on the number of selected samples and performs poorly under extreme compression. 
Graph-based distillation methods generally perform better by modeling structural dependencies. Among them, HGCond synthesizes data by matching gradients between the original and synthetic graphs, achieving the strongest performance and confirming the effectiveness of data synthesis. 
FreeHGC show limited improvement, as it relies on simple influence metrics instead of learnable feature generation.

Across different compression ratios, T2G maintains stable performance with minimal degradation.
In addition, the results show that a lower compression ratio does not necessarily lead to worse performance, which is consistent with previous findings in dataset distillation studies \cite{yu2023dataset}. This observation indicates that effective RDD can retain high predictive accuracy even under substantial compression.

In terms of task types, our proposed T2G effectively handles both regression and classification tasks across multiple databases and compression ratios.
Overall, these results highlight the effectiveness of T2G in preserving predictive power while achieving substantial data reduction.

Our evaluated databases exhibit diverse characteristics. For example, Airbnb contains highly imbalanced categorical columns, with imbalance ratio ranging from 2\% to 0.001\%. Despite these skewed distributions, our method addresses imbalance through independent tokenizers that assign each category a separate embedding. These embeddings are independently optimized on the original data, preventing majority categories from overwhelming minority representations.

In addition, the relation sparsity of our evaluated databases ranges from 0.1\% to 2\%. Despite this variation, our SBM module adaptively generates synthetic edges using the cluster-level sparsity defined in Eqs. (\ref{prob})–(\ref{eq:edge_gen}). By modeling connection probabilities between clusters rather than relying on global sparsity, it captures local interaction patterns and handles diverse sparsity effectively.

\begin{table}[t]
\renewcommand{\arraystretch}{1.1}
\centering
\caption{
The generalizability comparison of data reduction methods. Compression ratio $r=0.1\%$.
}
\label{tab:gene}
\resizebox{\linewidth}{!}{
\begin{tabular}{lccccccc}
\hline
\multicolumn{1}{l|}{\textbf{Models}} & \multicolumn{1}{l}{\textbf{SAGE}} & \multicolumn{1}{l}{\textbf{GIN}} & \multicolumn{1}{l}{\textbf{EDGE}} & \multicolumn{1}{l}{\textbf{HGT}} & \multicolumn{1}{l}{\textbf{HAN}} & \multicolumn{1}{l|}{\textbf{MLP}}   & \multicolumn{1}{l}{\textbf{Avg.}} \\ \hline
\multicolumn{8}{c}{\textbf{Rossmann $\downarrow$}}                                                                                                                                                                                                                                                      \\ \hline
\multicolumn{1}{l|}{Random}          & 352.9                                  & 318.3                            & 340.7                                & 348.9                            & 317.7                            & \multicolumn{1}{r|}{347.6}          & 337.7                             \\
\multicolumn{1}{l|}{Herding}         & 221.8                                  & 254.5                            & 265.8                                & 242.8                            & 245.4                            & \multicolumn{1}{r|}{241.7}          & 245.3                             \\
\multicolumn{1}{l|}{K-Center}        & 261.3                                  & 285.4                            & 296.8                                & 305.3                            & 276.3                            & \multicolumn{1}{r|}{300.3}          & 287.6                             \\
\multicolumn{1}{l|}{Coarsening}      & 236.0                                  & 255.9                            & 276.5                                & 257.3                            & 296.4                            & \multicolumn{1}{r|}{297.4}          & 269.9                             \\
\multicolumn{1}{l|}{FreeHGC}         & 219.2                                  & 246.2                            & 254.6                                & 246.8                            & 233.1                            & \multicolumn{1}{r|}{246.2}          & 241.0                             \\
\multicolumn{1}{l|}{HGCond}          & 213.1                                  & 214.6                            & 207.4                                & 247.1                            & 221.4                            & \multicolumn{1}{r|}{240.6}          & 224.0                             \\
\multicolumn{1}{l|}{T2G (Ours)}             & \textbf{192.8}                         & \textbf{196.2}                   & \textbf{190.9}                       & \textbf{213.7}                   & \textbf{216.5}                   & \multicolumn{1}{r|}{\textbf{221.3}} & \textbf{205.2}                    \\ \hline
\multicolumn{8}{c}{\textbf{Airbnb $\uparrow$}}                                                                                                                                                                                                                                                          \\ \hline
\multicolumn{1}{l|}{Random}          & 57.5                                   & 55.4                             & 55.7                                 & 56.1                             & 56.9                             & \multicolumn{1}{r|}{54.5}           & 56.0                              \\
\multicolumn{1}{l|}{Herding}         & 58.3                                   & 56.8                             & 57.9                                 & 57.3                             & 57.4                             & \multicolumn{1}{r|}{57.1}           & 57.5                              \\
\multicolumn{1}{l|}{K-Center}        & 57.9                                   & 56.1                             & 56.9                                 & 56.9                             & 57.2                             & \multicolumn{1}{r|}{57.1}           & 57.0                              \\
\multicolumn{1}{l|}{Coarsening}      & 59.0                                   & 57.3                             & 57.4                                 & 57.5                             & 57.7                             & \multicolumn{1}{r|}{56.8}           & 57.6                              \\
\multicolumn{1}{l|}{FreeHGC}         & 60.9                                   & 57.9                             & 58.6                                 & 58.0                             & 58.2                             & \multicolumn{1}{r|}{57.9}           & 58.6                              \\
\multicolumn{1}{l|}{HGCond}          & 62.6                                   & 62.7                             & 62.6                                 & 59.4                             & 62.5                             & \multicolumn{1}{r|}{62.8}           & 62.1                              \\
\multicolumn{1}{l|}{T2G (Ours)}             & \textbf{66.6}                          & \textbf{64.7}                    & \textbf{64.9}                        & \textbf{64.4}                    & \textbf{65.2}                    & \multicolumn{1}{r|}{\textbf{63.5}}  & \textbf{64.9}                     \\ \hline
\end{tabular}}
\end{table}

\subsection{Generalizability Comparison (Q2)}

The synthetic data is expected to generalize across diverse downstream model architectures. 
To thoroughly evaluate this property, we assess the performance of baseline methods under diverse experimental settings. 
We first conduct evaluations using a unified heterogeneous graph neural network framework~\cite{schlichtkrull2018modeling}, combined with different message passing paradigms, including GraphSAGE (SAGE)~\cite{hamilton2017inductive}, GIN~\cite{xu2018powerful}, and EdgeCNN (EDGE)~\cite{wang2019dynamic}. 
Furthermore, we examine two specialized heterogeneous graph models—HGT~\cite{hu2020heterogeneous} (transformer-based) and HAN~\cite{wang2019heterogeneous} (attention-based)—to further assess the adaptability of the synthetic data. 
In addition, we also include a simple MLP as a non-graph model to verify the generalization capability of the synthetic data beyond message passing architectures.

The detailed results are summarized in Table~\ref{tab:gene}. 
We observe that models trained on the synthetic data produced by T2G consistently achieve comparable performance across all architectures. 
Among these models, GraphSAGE yields the best performance due to its architectural alignment with the model used during distillation. 
In addition, the MLP, which discards the graph structure, still achieves comparable performance. This result suggests that the synthetic data effectively preserve the underlying structural information. 
When compared with other baselines, T2G significantly outperforms competing methods, demonstrating its strong ability to preserve predictive information and generalize effectively across diverse heterogeneous modeling paradigms.

\subsection{Ablation Study (Q3)}
\label{sec_ablation}

To evaluate the contribution of each component in T2G, we conduct an ablation study by disabling specific modules. 
As shown in Table~\ref{tab:ablation}, we first compare T2G without the SBM-based graph generation, where the graph structure is instead constructed by selecting the original edges following~\cite{gao2024heterogeneous, liang2025training}.  
Next, we remove the pseudo-label guidance in the feature distillation stage, denoted as ``{w/o}~$\mathcal{L}_{\text{pseudo}}$''.  
Finally, we entirely remove the pseudo-label in the process, referred to as ``{w/o pseudo-labels}''. 
In this setting, the pretraining stage is replaced by task-supervised training using task labels only, the SBM module is replaced by edge selection from the original database, and feature distillation relies solely on task labels.

When the SBM-based graph generation is replaced by sampling edges from the original REG, the relational connectivity among entities is not properly preserved, leading to structural information loss and a noticeable performance drop. 
In contrast, incorporating pseudo-label guidance during feature distillation allows each synthetic feature to be explicitly aligned with a coherent group of original entities. 
This divide-and-conquer mechanism enhances the representation consistency within each cluster and proves particularly beneficial for regression tasks, where capturing continuous label correlations is essential. 
Finally, when pseudo-label modeling is entirely removed from all stages of T2G, the performance degrades substantially, underscoring the pivotal role of pseudo-labels in guiding both structure generation and feature alignment. 

\begin{table}[t]
\setlength{\abovecaptionskip}{1pt}
\renewcommand{\arraystretch}{1.1}
\centering
\caption{Ablation study on three databases. The compression ratio is set to 0.1\%. The prediction tasks for Rossmann and Walmart are regression evaluated by MAE, while Airbnb is a classification task evaluated by AUC (\%).}
\resizebox{0.83\linewidth}{!}{
\begin{tabular}{lccc}
\Xhline{1.pt}
\textbf{Methods} & \textbf{Rossmann} $\downarrow$ & \textbf{Walmart} $\downarrow$ & \textbf{Airbnb} $\uparrow$ \\
\hline
w/o SBM           & 199.9$\pm$4.0 & 9334$\pm$218 & 65.0$\pm$0.6 \\
w/o $\mathcal{L}_{{pseudo}}$ & 194.5$\pm$3.5 & 9120$\pm$224 & 66.1$\pm$0.4 \\
w/o pseudo-labels & 207.4$\pm$4.3 & 9536$\pm$255 & 64.9$\pm$0.5 \\
Ours              & \textbf{192.8$\pm$4.1} & \textbf{8989$\pm$255} & \textbf{66.6$\pm$0.5} \\
\Xhline{1.pt}
\end{tabular}
}
\label{tab:ablation}
\end{table}

\subsection{Hyper-parameter Sensitivity Analysis (Q4)}
\label{sec_apphyper}
\begin{figure}[t]
\setlength{\abovecaptionskip}{1pt}
\centering
\includegraphics[width=0.85\linewidth]{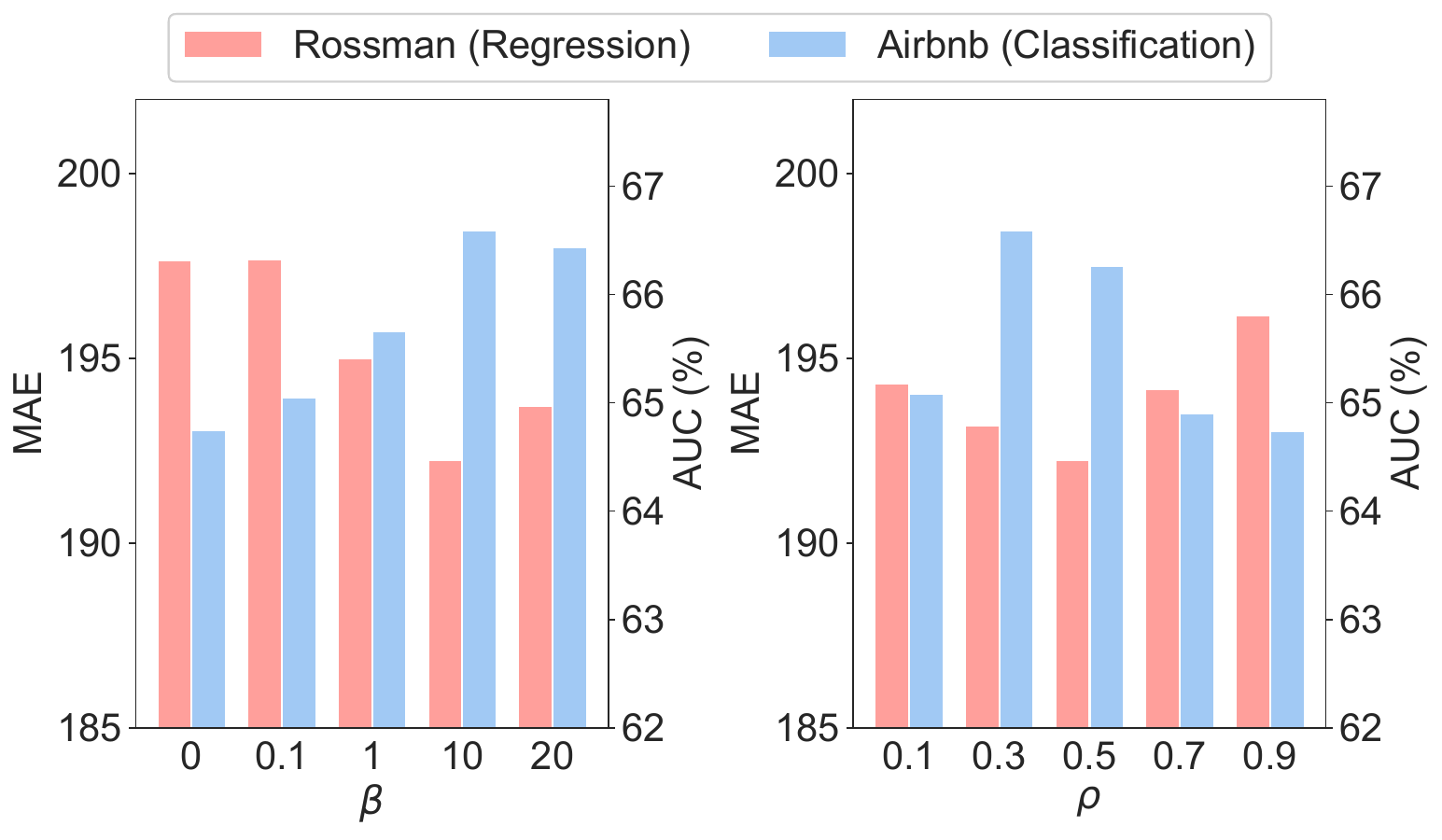}
\caption{Hyper-parameter analysis on Rossmann and Airbnb. The compression ratios are 0.1\%.}
\label{fig_hyper}
\end{figure}

Figure \ref{fig_hyper} presents the effect of $\beta$ and $\rho$ on both regression (Rossmann) and classification (Airbnb) tasks, where lower MAE and higher AUC indicate better performance.
The hyper-parameter $\beta$ controls the relative weight of the pseudo-label loss. As shown in the left sub-figure, increasing $\beta$ initially improves performance on both tasks by encouraging the synthetic features to align more closely with cluster-level semantics captured by the pseudo-labels. However, overly large $\beta$ values may overemphasize pseudo-labels and weaken overall supervision from task labels, leading to slight degradation.
For graph sparsity $\rho$ (right sub-figure), moderate sparsity yields the best performance. A smaller $\rho$ retains essential structural relations and avoids noise, consistent with findings in graph structure learning \cite{zhu2021survey} that sparser graphs often promote generalization. Yet, when $\rho$ becomes too small, insufficient edge information limits message propagation, causing performance to drop.

\section{Conclusion}
\label{sec_conclusion}
In this paper, we presented Table-to-Graph (T2G), the first framework for relational database distillation, which compresses large-scale relational databases into compact heterogeneous graphs while preserving predictive performance. T2G introduces a clustering-based pretraining objective that enables lightweight, modality-specific tokenization for multi-modal table attributes. Moreover, a SBM mechanism is employed to reconstruct heterogeneous graph structures guided by pseudo-labels.
To support both classification and regression tasks, T2G incorporates a KRR-based distillation objective that effectively transfers predictive knowledge into the synthetic features.
Extensive experiments on real-world relational databases show that T2G significantly reduces storage requirements while maintaining competitive performance. As an initial attempt at relational database distillation, T2G focuses on a single target table and task, leaving extensions to cross-table and cross-task settings for future work.

\begin{acks}
The Australian Research Council partially supports this work under the streams of Future Fellowship (Grant No. FT210100624), the Discovery Project (Grant No. DP240101108 and DP260100326), and the Linkage Project (Grant No. LP230200892 and LP240200546).
\end{acks}

\bibliographystyle{ACM-Reference-Format}
\balance
\bibliography{ref}

\appendix
\section{Appendix}

\subsection{Time Complexity}
\label{Tcom}

The overall pipeline of {T2G} comprises three main components: {pretraining}, {graph structure generation}, and {feature distillation}. We analyze the time complexity of each component separately.
For the original relational database, let $n_T$, $e_T$, $d_{{num}}$, and $d_{{cat}}$ denote the average number of entities, links, numerical columns, and categorical columns per table, respectively. The numbers of table and link types are represented by $n$ and $m$. The average number of entities in the synthetic REG is denoted by $n'_T$, and the token and embedding dimensions are $d_{{token}}$ and $d$.
\begin{itemize}[leftmargin=*]
\item Pretraining. This stage includes column tokenization, $K$-layer graph convolution, and online clustering. 
The time complexities of these submodules are: $\mathcal{O}\!\left(n n_T d_{\text{token}} (d_{\text{num}} + d_{\text{cat}})\right)$, $\mathcal{O}\!\left(m K e_T d\right)$, and $\mathcal{O}\!\left(nn'_T n_T d t\right)$, where $t$ denotes the number of iterations in the clustering.

\item Graph Structure Generation. The generation process includes SBM estimation and adjacency sampling, which require  $\mathcal{O}(me_T)$ and $\mathcal{O}(mn_T'^2)$, respectively.

\item Feature Distillation. This stage involves $K$-step message passing in the HGNN and the closed-form solution of KRR. 
Their respective complexities are:
$\mathcal{O}\!\left(m K e_T d\right)$ and
$\mathcal{O}\!\left(n_T'^3 + n_T d D\right)$,
where $D$ denotes the dimension of the predictor head (i.e., the number of categories for classification or $D=1$ for regression task).
\end{itemize}
In a nutshell, the time complexity of {T2G} is dominated by the pretraining and feature distillation stages. 
However, since $n'_T \ll n_T$ and efficient acceleration techniques (e.g., FAISS~\cite{douze2024faiss} and neighbor sampling in HGNNs) can be applied, the overall computational cost remains well controlled, ensuring scalability to large-scale RDBs.

\subsection{Storage Efficiency Analysis (Q5)}
\label{sec_distilldata}
Table~\ref{tab:storage} summarizes the storage comparison between the original databases and the synthetic data generated by T2G.
Across all three benchmarks, T2G achieves remarkable data compression while preserving essential predictive information.
For instance, Rossmann’s synthetic data retains only 0.02\% of the original rows in the training set yet reduces the storage size from 35.0 MB to 61.1 KB, achieving over 500× compression.

A small portion of the total storage comes from the tokenizers, which encode column-specific modality representations.
Although tokenizers add only a few kilobytes overall, they are crucial for enabling consistent encoding of numerical and categorical attributes.
Notably, Airbnb exhibits a relatively larger tokenizer size (23.0 KB) compared to Rossmann (5.0 KB) and Walmart (8.7 KB).
This increase is attributed to Airbnb’s higher column dimensionality and richer categorical diversity, which require more embedding parameters to represent diverse attributes.
Nevertheless, the tokenizer storage remains minimal compared to the original database size, contributing less than 0.01\% of the total storage.


\begin{table}[th]
\setlength{\abovecaptionskip}{1pt}
\renewcommand{\arraystretch}{1.1}
\centering
\caption{Comparison of size and storage. $r$ denotes the compression ratio, defined as the proportion of entities in the synthetic data relative to those in the original \textit{training set}.}
\label{tab:storage}
\resizebox{\linewidth}{!}{%
\begin{tabular}{l|cc|cc|cc}
\Xhline{1.pt}
& \multicolumn{2}{c|}{\textbf{Rossmann}} & \multicolumn{2}{c|}{\textbf{Walmart}} & \multicolumn{2}{c}{\textbf{Airbnb}} \\ \cline{2-7}
& Original & Ours & Original & Ours & Original & Ours \\
& Database & ($r$=0.02\%) & Database & ($r$=0.1\%) & Database & ($r$=0.02\%) \\ 
\hline
\#Rows & 1,015,159 & 147 & 310,707 & 192 & 1,020,723 & 119 \\ 
\hline
Data & 35.0M & 59.1KB & 13.5MB & 80.8KB & 350.0MB & 46.8KB \\
Tokenizers & N/A & 5.0KB & N/A & 8.7KB & N/A & 23.0KB \\
\hline
\textbf{Total Storage} & \textbf{35.0M} & \textbf{61.1KB} & \textbf{13.5MB} & \textbf{89.5KB} & \textbf{350.0MB} & \textbf{69.8KB} \\
\Xhline{1.pt}
\end{tabular}}
\end{table}

\end{document}